\documentclass[journal]{IEEEtran}
\usepackage{amsmath,amssymb,}
\usepackage{setspace}
\usepackage{graphicx}
\usepackage{tikz}
\usepackage{standalone} 
\usepackage{xcolor} 
\usepackage{color}
\usepackage{lipsum,adjustbox}
\usetikzlibrary{decorations}
\usepackage{pgfplots,siunitx}
\usetikzlibrary{decorations.pathreplacing}
\usetikzlibrary{shapes,arrows,shadows}
\usetikzlibrary{calc,positioning}
\usetikzlibrary{fit,shapes}
\usepackage{tikz,tkz-euclide}
\usepackage{balance}
\usepackage{cite}
\usepackage[pdftex,colorlinks=true,bookmarks=false,citecolor=blue,urlcolor=blue]{hyperref}

\newcommand{\EmaxCCDM}[0]{{E}^{\circ}}
\newcommand{\EmaxESS}[0]{{E}^{\bullet}}

\def\calX{\mathcal{X}} 
\def\calS{\mathcal{S}}
\def\calA{\mathcal{A}} 

\pgfplotsset{compat=1.3}
\newtheorem{example}{Example}
\usetikzlibrary{calc}

\begin{document}
\title{Introducing Enumerative Sphere Shaping for Optical Communication Systems with Short Blocklengths}
\author{Abdelkerim Amari, \textit{Member, IEEE}, Sebastiaan Goossens, \textit{Student Member, IEEE},  Yunus Can G\"{u}ltekin, \textit{Student Member, IEEE},  Olga Vassilieva, \textit{Senior Member, IEEE}, Inwoong Kim, \textit{ Member, IEEE}, Tadashi Ikeuchi, \textit{Member, IEEE}, Chigo Okonkwo, \textit{Senior Member, IEEE}, Frans M. J. Willems, \textit{Fellow, IEEE}, and Alex Alvarado, \textit{Senior Member, IEEE} 
\thanks{A. Amari, Y. C. G\"{u}ltekin,  and F. M. J. Willems, and A. Alvarado are with the Information and Communication Theory Lab, Signal Processing Systems Group, Department of Electrical Engineering, Eindhoven University of Technology, Eindhoven 5600 MB, The Netherlands.
Email: a.amari@tue.nl.}
\thanks{S. Goossens and C. Okonkwo are with the Institute for Photonic Integration,
Eindhoven University of Technology, Eindhoven 5600 MB, The Netherlands.}

\thanks{O. Vassilieva, I. Kim, and T. Ikeuchi are with Fujitsu Laboratories of America, Inc., Richardson, TX 75082 USA.}
}

\maketitle

\begin{abstract}
Probabilistic shaping based on constant composition distribution matching (CCDM) has received considerable attention as a way to increase the capacity of fiber optical communication systems. CCDM suffers from significant rate loss at short blocklengths and requires long blocklengths to achieve high shaping gain, which makes its implementation very challenging. In this paper, we propose to use enumerative sphere shaping (ESS) and investigate its performance for the nonlinear fiber optical channel. ESS has lower rate loss than CCDM at the same shaping rate, which makes it a suitable candidate to be implemented in real-time high-speed optical systems. In this paper, we first show that finite blocklength ESS and CCDM exhibit higher effective signal-to-noise ratio than  their infinite blocklength counterparts. These results show that for the nonlinear fiber optical channel, large blocklengths should be avoided. We then show that for a $400$ Gb/s dual-polarization 64-QAM WDM transmission system, ESS with short blocklengths outperforms both uniform signaling and CCDM. Gains in terms of both bit-metric decoding rate and bit-error rate are presented. ESS with a blocklength of $200$ is shown to provide an extension reach of about $200$~km in comparison with CCDM with the same blocklength. The obtained reach increase of ESS with a blocklength of $200$ over uniform signaling is approximately $450$~km (approximately $19\%$).
\end{abstract}
\IEEEpeerreviewmaketitle

\section{Introduction}
\IEEEPARstart{O}{ptical} communication systems have evolved to ensure the growth of Internet traffic, which has exponentially increased in recent years due to multiple modern bandwidth-hungry applications. Fiber optical systems are approaching their capacity limits and an optimal exploitation of the installed network resources is highly desirable to keep up with the traffic demands \cite{bayvel}. In this context, several approaches have been investigated in the literature to increase the spectral efficiency. One of the most popular techniques is fiber nonlinearity mitigation, which can be implemented using different approaches \cite{winzer,am1}. The main drawback of fiber nonlinearity compensation is its high complexity, which makes its real-time implementation very challenging. Another way to increase the spectral efficiency is by employing sophisticated forward error correcting (FEC) combined with high-order modulation formats, a combination known as coded modulation \cite{Szczecinski,Alvarado2015_JLT}. Coded modulation can be taken one step further via constellation shaping, a technique that dates back to the early 1990s \cite{Calderbank90,Forney92,Kschischang93}.

Constellation shaping has recently received a lot of attention in the fiber optical communications community. Two types of constellation shaping have been proposed: geometric shaping \cite{gs0,gs1,gs2,gs3,gs4}, and probabilistic shaping \cite{ps0,ps1,ps2,ps3,ps4,ps5,ps6,mp,ps7,ps8, mp2, dd,R3}. Hybrid probabilistic and geometric shaping have been also investigated as in \cite{gsps1,gsps2}. While probabilistic shaping consists of using uniformly spaced constellation points with different probabilities, geometric shaping uses nonuniformly spaced constellation points with same probabilities. In the additive white Gaussian noise (AWGN) channel, both techniques can provide gains up to $1.53$ dB. This so-called ``ultimate shaping gain'' is achieved at infinite blocklengths (for probabilistic shaping) or infinitely dense constellations (for geometrical shaping). The ultimate shaping gain can in principle be exceeded for the nonlinear fiber optical channel, as shown in \cite{DarISIT14}. Probabilistic shaping has been shown to give larger gains in comparison with geometric shaping, especially when bit-metric decoding is used \cite[Ch.~4]{Szczecinski}, \cite{pg}.  

 Probabilistic shaping has been widely investigated in the context of optical communication systems not only because of its shaping gain but also because it provides rate adaptivity for systems with fixed FEC \cite{ps5}. Distribution matching (DM) is considered as a key technique for the implementation of the shaping and deshaping of probabilistically shaped sequences. So far, constant composition DM (CCDM) \cite{ps0} with long blocklengths is the preferred alternative to achieve the Maxwell-Boltzmann (MB) distribution, which maximizes the shaping gain in AWGN channel. However, it has been observed in \cite{ps2} that CCDM-based probabilistic shaping has lower tolerance to the fiber nonlinearity in comparison with uniform distribution. Nevertheless, the combination of shaping gain and nonlinear penalty results in an overall increase of performance with respect to uniform signaling \cite{ps2}. The total gain, however, is lower than for the AWGN channel due to the nonlinear penalty. CCDM with long blocklengths is also very difficult to implement in high-speed communications because it is based on \emph{sequential} arithmetic coding. A way to decrease the blocklength by reducing the rate loss, known as multiset-partition DM, has been recently presented in \cite{mp, mp2}. Sphere shaping implementations based on shell mapping and look-up table DM,  have been also proposed \cite{ps8,R1,R3} to improve the performance in optical systems.
 
 In this paper, we propose to use enumerative sphere shaping (ESS) 
 as a means to realize sphere shaping in the context of optical communication systems\footnote{During the revision process of this paper, we have published two conference papers in which ESS is evaluated numerically~\cite{KarimECOC} and experimentally~\cite{SebastianPDP2019} for optical communication systems.}, and investigate its performance in comparison with CCDM. 
 Sphere shaping can be implemented using different ways such as: ESS, which has been introduced in \cite{willems}, shell mapping, and the approach in \cite{Laroia}. Recently, after the proposal of probabilistic amplitude shaping (PAS) scheme \cite{ps5}, ESS has been reconsidered for wireless communications as the amplitude shaper in PAS structure \cite{gultekin2,gultekin4}. ESS considers amplitude sequences satisfying a maximum-energy, i.e., sphere, constraint and aims to close the shaping gap by improving the energy-efficiency rather that directly aiming to obtain the capacity-achieving distribution as CCDM. However, sphere shaping achieves the MB distribution at infinite shaping blocklength \cite{gultekin1}. In addition, there is also an exact relation between the MB distribution and sphere shaping for finite blocklength: for a given shaping rate, sphere shaping minimizes the informational divergence between the distribution at the DM output and an MB distribution \cite{R1}.
 
In this paper, we compare the performance of ESS and CCDM through numerical simulations. To the best of our knowledge, this is the first time such comparison is made in the context of the nonlinear fiber optical channel. Both shaping approaches improve the performance in comparison with uniform signaling at long blocklengths. At short blocklengths, ESS is shown to outperform CCDM due to its lower rate loss. In addition, we show that short blocklengths ESS and also CCDM provide \emph{higher} effective signal-to-noise ratio (SNR) than uniform 64-QAM. This is in contrast to the long blocklengths case, where probabilistic shaping causes an effective SNR reduction. 
 
The key observation in this paper is that short blocklengths ESS should be the preferred alternative for shaping in the nonlinear fiber optical channel. This is due to a balance between linear and nonlinear gains, low rate loss, and high effective SNR. Bit-metric decoding (BMD) rate results confirm that the best performance is obtained for short blocklengths ESS. The BMD rate of short blocklength ESS is shown to be higher than long blocklengths ESS, short and long blocklengths CCDM, and uniform signaling. For a $400$ Gb/s dual-polarization 64-QAM WDM transmission system, the effecive SNR and BMD rate gains are translated to considerable transmission reach increases. For ESS with a blocklength of $200$, gains of about $100$ km, $200$ km, and $450$ km are reported  in comparison with CCDM with  blocklength of $3600$, CCDM with  blocklength of  $200$ and uniform, resp. End-to-end results including shaping and FEC are also presented in this paper.

The remainder of the paper is organized as follow. The system model and signaling schemes are discussed in Sec.~\ref{sec:SHPM}. The working principles of CCDM and ESS techniques are reviewed in Sec.~\ref{sec:ccdm_ess}. The simulation setup, performance metrics, and  numerical results are presented in Secs.~\ref{sec:simr.1} and \ref{sec:simr.2}. Conclusions are drawn in Sec.~\ref{sec:con}.

\section{System Model and Signaling Schemes}\label{sec:SHPM}
 
\begin{figure*}[t]
\centering		
\includegraphics{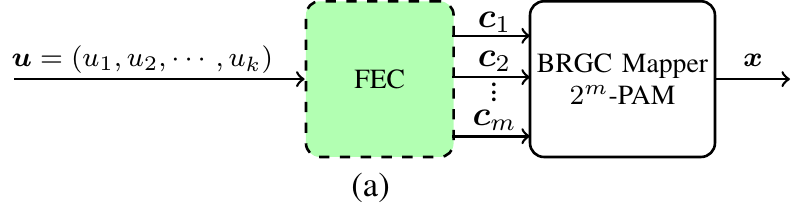}
\includegraphics{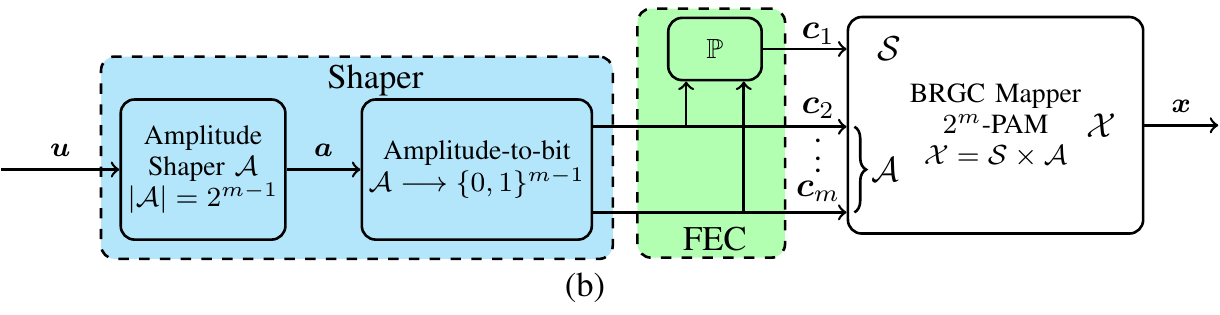}
\includegraphics{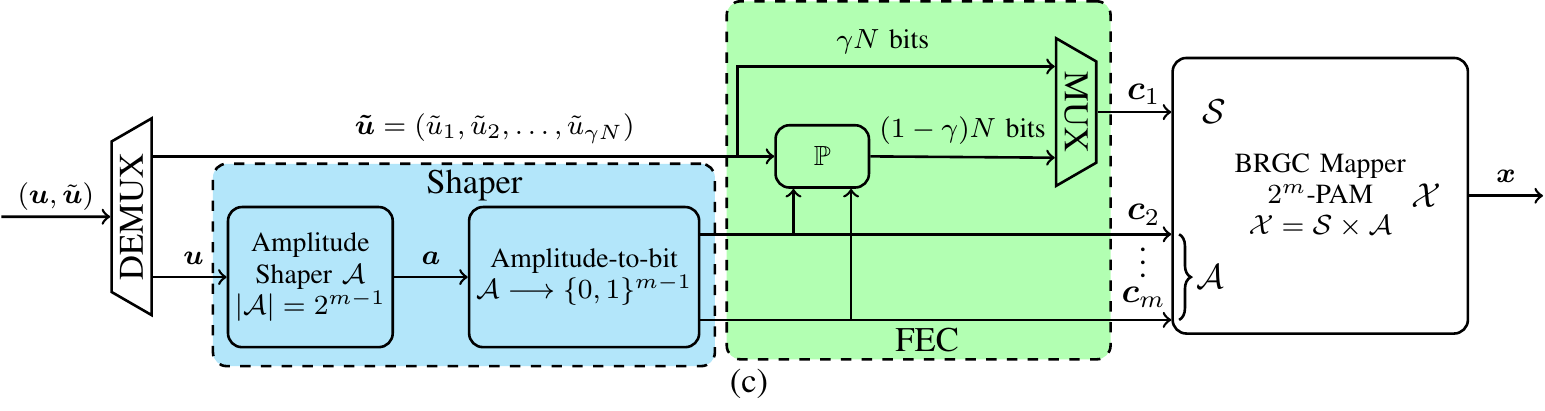}
\caption{Signaling alternatives: (a) uniform signaling, (b) PAS architecture with FEC rate $R_c = (m-1)/m$, and (c) PAS architecture with FEC rate $R_c > (m-1)/m$.}
\label{fig:PASblockdiag}
\end{figure*}

In this work, we consider the PAS scheme as an alternative to uniform signaling schemes.
We assume transmission using polarization multiplexed (PM) square QAM constellations. The four-dimensional signal space (PM-QAM signals) is generated by the Cartesian product of four identical $2^m$-ary pulse amplitude modulation (PAM) constellations using the same binary labeling. Without loss of generality, we consider the underlying $2^m$-PAM constellation. The alphabet (constellation) has the form $\calX = \{\pm 1, \pm 3,\cdots, \pm 2^m-1 \}$ where $m \geq 2$.
These alphabets can be factorized as $\calX = \calS \times \calA$, where $\calS = \{-1, +1\}$ and $\calA = \{1, 3,\cdots, 2^m-1\}$. We refer to $\calS$ and $\calA$ the sign and amplitude alphabets, resp., where $|\calS|=2$ and $|\calA|=2^{m-1}$.

One requirement of the PAS scheme is that the binary label of the PAM constellation can be separated into two parts: a sign bit and $m-1$ amplitude bits. In this paper we consider the binary reflected Gray coded (BRGC) \cite{Gray53}, which due to its recursive construction \cite[Sec.~IV]{Agrell04} satisfies this condition. The BRGC is also known to be asymptotically the best binary labeling in terms of maximizing the generalized mutual information (GMI) \cite[Sec.~IV]{Alvarado12b}.

\subsection{Uniform Signaling}

In uniform signaling, a $k$-bit uniform information sequence $\boldsymbol{u}=(u_1,u_2,\ldots,u_k)$ is encoded by a rate $R_c = k/n_c$ FEC code.\footnote{Throughout this paper, boldface letters $\boldsymbol{x}$ denote vectors and blackboard letters $\mathbb{X}$ denote matrices.} The $n_c$-bit coded sequence $\boldsymbol{c}=(c_1,c_2,\ldots,c_n)$ is then divided into length-$m$ binary vectors $\boldsymbol{b}=(b_1,b_2,\ldots,b_m)$, each of which is mapped to a channel input $x$ via the BRGC. The sequence of symbols transmitted through the channel is $\boldsymbol{x}=(x_1,x_2,\ldots,x_N)$, where $N$ is the number of transmitted symbols. The information rate of this scheme is $R=k/N=mR_c$~[bits/1D-sym], or equivalently, $2mR_c$~[bits/2D-sym]. This scheme is shown in Fig.~\ref{fig:PASblockdiag}~(a).

\subsection{Nonuniform Signaling}

The PAS architecture shown in Fig.~\ref{fig:PASblockdiag}~(b) is a reverse concatenation structure, where the shaping operation precedes FEC encoding.
First, an amplitude shaping block transforms the uniform information sequence $\boldsymbol{u}$ into a shaped amplitude sequence $\boldsymbol{a}=(a_1,a_2,\ldots,a_N)$, where $a_n\in\calA$ with $n=1,2,\ldots,N$. The sequences $\boldsymbol{a}$ are chosen to satisfy a predefined condition. The rate of this shaper, referred to as the shaping rate, is $R_s=k/N$~[bits/amp]. This conversion is an invertible mapping which can be realized by various shaping approaches (e.g., CCDM or ESS). Next, the amplitudes $\boldsymbol{a}$ are mapped into bits using the last $m-1$ bits of the BRGC. This is shown in Fig.~\ref{fig:PASblockdiag}~(b) as amplitude-to-bit conversion, which generates $N(m-1)$ nonuniform bits. 

In the simplest form of the PAS architecture, the $N(m-1)$ nonuniform bits are fed to a rate $R_c = (m-1)/m$ systematic FEC encoder, which generates $Nm$ code bits and $N$ parity bits. As shown in Fig.~\ref{fig:PASblockdiag}~(b), this can accomplished using a parity-check matrix $\mathbb{P}$ of dimensions $(m-1)N\times N$. The $N$ parity bits are mapped to the first bit of the $2^m$-PAM mapper, and thus, they select the signs (they are mapped to $\calS$). These are called sign bits. The $N(m-1)$ nonuniform bits are used as bits $2$ to $m$ the $2^m$-PAM mapper, and thus, they are referred to as the shaping bits. Finally, the sequence $\boldsymbol{x}\subset S^N\times A^N$ is transmitted over the channel. The information rate of this scheme is $R=R_s=k/N$~[bits/1D-sym].

The simplest form of the PAS scheme in Fig.~\ref{fig:PASblockdiag}~(b) assumes a FEC rate of $R_c = (m-1)/m$. Lower FEC rates can be used if the bit-level shaping schemes in \cite{Steiner1,gultekin5} are considered. Higher FEC code rates ($R_c > (m-1)/m$) can be used via a straightforward modification of the PAS architecture. This is shown in Fig.~\ref{fig:PASblockdiag}~(c). The FEC rate in this case is assumed to be $R_c = (m-1+\gamma)/m$, where $0<\gamma<1$ represents additional information bits. As shown in Fig.~\ref{fig:PASblockdiag}~(c), the information sequence is of length $k+\gamma N$. In the lower branch, $k$ bits are shaped in the same way as in Fig.~\ref{fig:PASblockdiag}~(b). In the upper branch, $\gamma N$ information bits are used together with $(m-1)N$ shaped bits to generate $(1-\gamma)N$ parity bits. These $(1-\gamma)N$ parity bits are multiplexed with the uniform $\gamma N$ bits to create the vector $\boldsymbol{c}_1$ of $N$ sign bits. 

In the modified architecture in Fig.~\ref{fig:PASblockdiag}~(c), the parity-check matrix $\mathbb{P}$ is of dimensions $(m-1+\gamma)N\times N$. The information rate of this scheme is $R=(k+\gamma N)/N=R_s + \gamma$~[bits/1D-sym]. Note that the case $\gamma=0$ corresponds to  the scheme in Fig.~\ref{fig:PASblockdiag}~(b), while $\gamma=1$ corresponds to uncoded transmission.

\begin{example}\label{example1}
Consider the uniform signaling scheme which combines a rate $R_c=3/4$  FEC code with $8$-PAM ($m=3$). The transmission rate for this scheme is $R=R_cm=2.25$~[bits/1D-sym]. To achieve the same rate with a shaped 8-PAM constellation, we need to somehow compensate for the decrease in rate caused by shaping. Thus, we use a higher rate FEC code, more precisely $R_c=4/5$, which leads to $\gamma=0.4$. Combining this code with a shaping rate $R_s=1.85$~bits/amp shaper, we obtain the same rate, i.e., $R=R_s+\gamma=2.25$~[bits/1D-sym]. 
\end{example}

\section{Amplitude Shaping: Constant Composition and Sphere Coding}\label{sec:ccdm_ess}

A nonuniform channel input distribution can be obtained by changing the bounding geometry of the multidimensional signal space. Consider for example the transmission of $N$ amplitudes $\boldsymbol{a}=(a_1,a_2,\ldots,a_N)$, where each amplitude is taken from a set $\mathcal{A}$. If no coding is applied, the transmitted codewords are all the $|\mathcal{A}|^N$ $N$-dimensional vectors from the set $\mathcal{A}^N$. If FEC is applied, some sequences are never transmitted, and thus, the codewords belong to a subset of $\mathcal{A}^N$. Probabilistic shaping can be interpreted as the process of selecting the subset of \emph{allowed} sequences from $\mathcal{A}^N$. In what follows, we describe two approaches to do this: constant composition and sphere coding.

\subsection{Constant Composition Coding}\label{sec:ccdm}

CCDM consists in generating probabilistically shaped sequences, with a given probability distribution, from uniformly distributed bits. It is inspired by arithmetic coding for data compression, in which the sequences are represented by intervals \cite{ps0}. The performance of CCDM in terms of shaping gain is optimum at infinite blocklengths $N\rightarrow\infty$. In this regime, any desired distribution can be achieved, including the MB distribution, which is optimal for the AWGN channel. However, CCDM suffers from high rate loss at short blocklength \cite{mp,gultekin1}.

CCDM only allows sequences $\calA^\circ \subset \calA^N$ that satisfy a particular composition. A sequence $\boldsymbol{a}$ is valid only if out of the $N$ amplitudes, the sequence contains exactly $n_1$ amplitudes one, $n_2$ amplitudes three, $n_3$ amplitudes five, etc. The composition is therefore $(n_1,n_2,\ldots,n_{2^m-1})$. By definition, all the sequences generated by CCDM have the same per-codeword energy $\EmaxCCDM$, which depends on the targeted composition. 

\begin{example}[Geometry of Constant Composition Coding]
The geometrical interpretation of CCDM is therefore sequences on an $N$-dimensional sphere of radius $\sqrt{\EmaxCCDM}$. This is schematically shown in Fig.~\ref{fig:3}~(a), where a 2D projection is presented. The fact that CCDM does not fully cover the sphere comes from the fact that not all sequences on the shell of the $N$-dimensional sphere of radius $\sqrt{\EmaxCCDM}$ satisfy the composition constraint.
\end{example}

\subsection{Sphere Coding}\label{sec:ESS}

Sphere coding is the process of constraining the codewords to be selected from within an $N$-sphere. In this case, the induced distribution in one real dimension converges to a MB distribution as $N\rightarrow\infty$ (or to a Gaussian distribution if the set $\calA$ has a continuous support). 

The idea of sphere shaping is to utilize the set of bounded-energy (i.e., spherically-constrained) amplitude sequences
\begin{align}\label{:}
\calA^\bullet=\left\lbrace a_1,a_2,\ldots,a_N \Biggl| \sum_{n=1}^{N} a_n^2 \leqslant \EmaxESS \right\rbrace \subset \mathcal{A}^N,
\end{align}
where $\EmaxESS$ is a real constant. The value of $\EmaxESS$ determines the ``amount of shaping'': the smaller the value of $\EmaxESS$, the more shaped the signal will be. Note that if the alphabet $\calA$ is discrete, the energy levels will be quantized.

\begin{figure}[t]
\centering	
\resizebox{\columnwidth}{!}{ \includegraphics{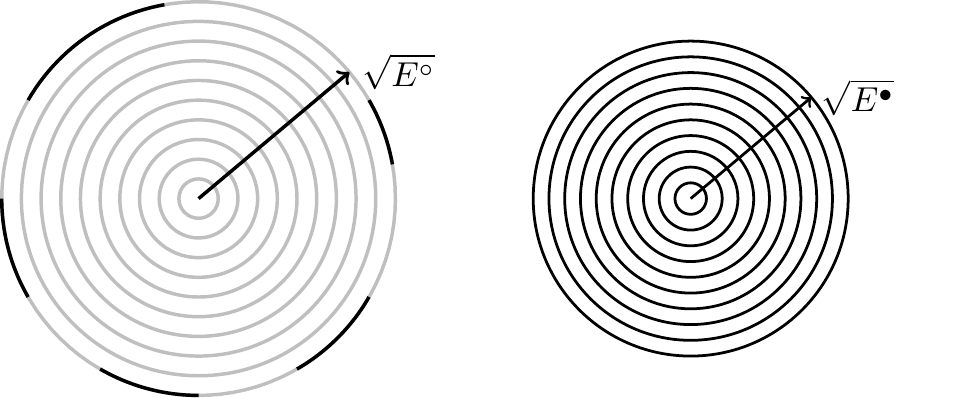} }
\caption{Geometric representation in 2D of shaping schemes: (a) Constant composition, and (b) Sphere shaping.}
\label{fig:3}
\end{figure}

\begin{example}[Geometry of Sphere Coding]
The set $\calA^\bullet$ in \eqref{:} consists of all amplitude sequences having an energy no greater than $\sqrt{\EmaxESS}$. Geometrically, this corresponds to all points inside or on the surface of an $n$-sphere of radius $\sqrt{\EmaxESS}$ are employed. This is shown schematically in Fig.~\ref{fig:3}~(b). This figure shows schematically that sphere coding can transmit more sequences (i.e., $|\calA^\bullet|>|\calA^\circ|$). Furthermore, the energy of the inner shells in $\calA^\bullet$ is smaller, and thus, the spheres in Fig.~\ref{fig:3}~(b) are smaller than those in Fig.~\ref{fig:3}~(a) (energy can be saved via sphere shaping).
\end{example}

Due to the sphere hardening, all sequences will concentrate near the surface of the $N$-sphere as $N\rightarrow\infty$ which means that CCDM and sphere shaping are asymptotically equivalent for large $N$. However, for finite blocklengths, there is a rate loss of CCDM with respect to sphere shaping. The rate loss is defined as \cite{gultekin1}
\begin{align}\label{eq:rloss}
R_L \triangleq H(P_A) - \frac{k}{N} = H(P_A) - R_s \left[\text{bits}/\text{amp} \right],
\end{align}
where $P_A$ is the targeted probability distribution and $H(P_A)$ is the entropy in bits/amp. 

\begin{figure}[t]
\centering		
\resizebox{\columnwidth}{!}{ \includegraphics{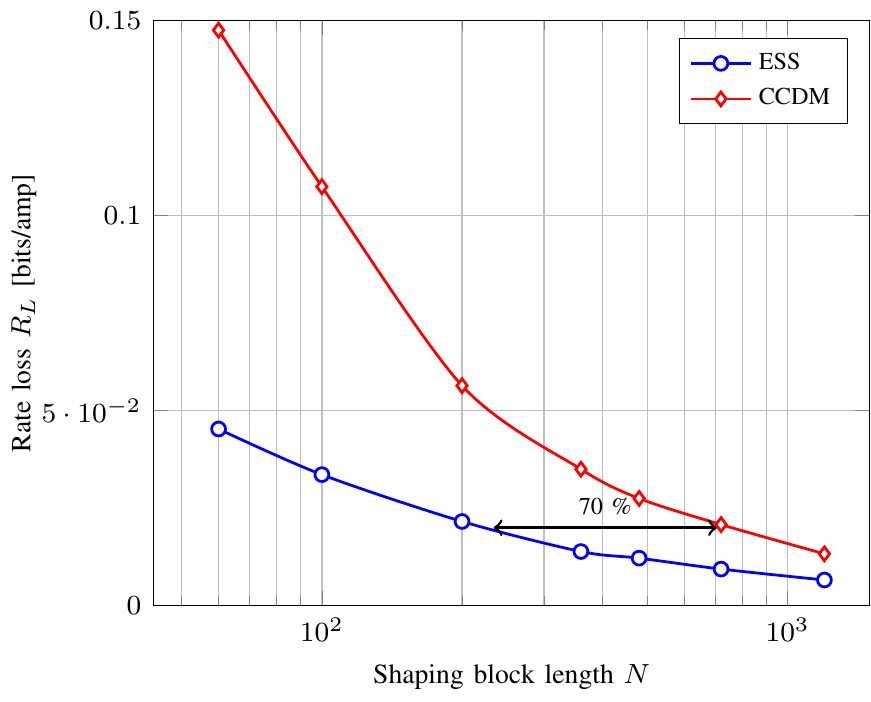} }
\caption{Rate loss vs. shaping blocklength for ESS and CCDM.}	
\label{fig:2}
\end{figure}

\begin{example}[Rate Losses]
The rate losses of CCDM and ESS are shown in Fig.~\ref{fig:2}.  At a rate loss of $0.02$ bits/amp, ESS reduces the shaping blocklength by about $70\%$ when used instead of CCDM. Similar results for an entropy of 1.75~bits/amp. were presented in \cite[Fig.~2]{gultekin4}.
\end{example}

\subsection{Enumerative Sphere Shaping}\label{sec:ESS}

Multiple algorithms have been proposed to realize sphere shaping by indexing the  bounded-energy amplitude sequences \cite{ps8,R1,R3,Laroia,willems}.
In this work, we use the enumerative approach introduced in \cite{willems}, and named enumerative sphere shaping (ESS) in \cite{gultekin1,gultekin2,gultekin4}.

ESS starts from the assumption that these sequences can be ordered lexicographically.
To create this ordering, a bounded-energy amplitude trellis is constructed, as shown in the following example.

\begin{example}[ESS Trellis]\label{ess.example}
Fig.~\ref{fig:4} shows a bounded-energy amplitude trellis with $N=4$, $\mathcal A=\{1, 3, 5,7\}$ and $\EmaxESS=60$. In this trellis, each state represents an energy level, which is indicated using black numbers.
Each branch designates an amplitude from $\calA$. 
Each path, starting from the zero-energy state (i.e., bottom left) and ending in one of the states in the final stage, represents an amplitude sequence. 
These sequences are composed of the corresponding amplitude values (different colors) of the branches of their path. 
The energy level at stage $n$ that a path passes through, is the accumulated energy of the sequence over the first $n$ components. 

\begin{figure}[t]
\centering		
\resizebox{\columnwidth}{!}{ \includegraphics{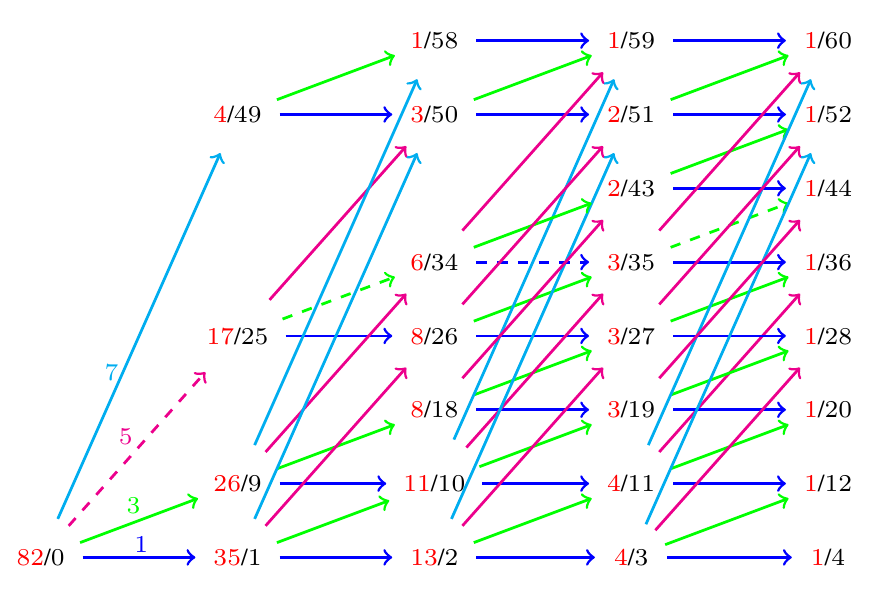} }
\caption{Enumerative trellis for $N=4$, $\calA = \{1, 3, 5, 7\}$ and $\EmaxESS=60$. 
Each state shows the number of unique paths advancing from that state to one of the states in the final stage (red numbers), and the accumulated energy of the paths arriving to that state (black numbers).}
\label{fig:4}
\end{figure}
\end{example}

A key number for the enumerative shaping and deshaping algorithms is $T_n^e$ which is the number  of total paths advancing from a state of energy $e$ in stage $n$ to one of the final states. 
$T_0^0$ is therefore the total number of sequences represented in the trellis, i.e., $T_0^0=|\calA^\bullet|$, which is $82$ in Example~\ref{ess.example}, see Fig.~\ref{fig:4}. 
The rate of the shaper that indexes sequences from $\calA^\bullet$ can then be expressed as 
\begin{align}\label{eq:shapingrate}
R_s =\frac{\log_2T_0^0}{N} \left[ \mbox{bits/amp} \right].
\end{align}
The values of $T_n^e$ (shown with red in Fig.~\ref{fig:4}), can be calculated in a recursive manner as
\begin{equation}
T_n^e \triangleq 
\sum_{a \in \calA} T_{n+1}^{e+a^2}, \label{trellisCons}
\end{equation}
where the states in the last stage are initialized with ones, hence
\begin{equation}
T_N^e = 
\begin{cases}
 1, & \text{if $e = N,N+8,N+16,\ldots,\EmaxESS$}\} \\
 0, & \text{otherwise}.
\end{cases}
\label{trellisEq}
\end{equation}

Finding the index $i$ of a sequence $\boldsymbol{a}=(a_1,a_2,\ldots ,a_N)$ is equivalent to count the number of sequences which are lexicographically smaller than $\boldsymbol{a}$. 
This can be implemented by considering the path in the trellis corresponding to $\boldsymbol{a}$.
Following this path, at stage $n$ for $n = 0, 1,\cdots, N-1$, we accumulate the number of sequences which have their first $n$ elements identical to $\boldsymbol{a}$ and are lexicographically smaller than $\boldsymbol{a}$.
The procedure starts in the zero-energy state.
All sequences which start with an amplitude $a<a_1$ have a smaller index than $\boldsymbol{a}$.
Thus we sum the corresponding $T_1^{a^2}$ values and arrive at the state of energy $a_1^2$ in the first stage.
At this point, all sequences which start with $a_1$ and continue with an amplitude $a<a_2$ have a smaller index than $\boldsymbol{a}$.
Thus we add the corresponding $T_2^{a_1^2 + a^2}$ values to our accumulating sum and arrive at the state of energy $a_1^2 + a_2^2$ in the second stage.
Repeating this procedure recursively, we arrive at the state of energy $\sum_i a_i^2$ at final stage, and the accumulated sum is the index $i(\boldsymbol{a})$ of our sequence $\boldsymbol{a}$.
This leads to Cover's indexing formula, see~\cite{Cover1973},
\begin{equation}
i(\boldsymbol{a}) = \sum_{n=1}^N \sum_{{b\in\calA: b<a_n}} T_n^{b^2 + \sum_{m=1}^{n-1} a_m^2}. \label{eq:essindex}
\end{equation}
The following example demonstrates \eqref{eq:essindex} based on the trellis in Fig.~\ref{fig:4}. 

\begin{example}[ESS Indexing]\label{ess.indexample}
Consider the sequence $\boldsymbol{a} = (5,3,1,3)$, which is represented by a path passing through energy levels $\boldsymbol{e} = (0, 25, 34, 35,44)$ in stages $n =0, 1, 2, 3,4$, respectively. 
This path is shown with dashed lines in Fig.~\ref{fig:4}.
At first, we count the number of sequences which start either with a $1$ or a $3$ since all these will have a smaller index than $\boldsymbol{a}$.
Thus we add the numbers $T_1^1=35$ and $T_1^9=26$ to get $61$, and arrive at the node of energy $a_1^2 = 25$ in the first stage.
Then we need to add the number of sequences which start with $(5,1)$ since all these will have a smaller index than $\boldsymbol{a}$.
Therefore we add $T_2^{26}=8$ to our accumulating sum to get $69$.
Our sequence have a $1$ in the third position which makes it lexicographically the first, given that the first two elements are fixed.
Thus we add nothing in this stage.
Finally, we need to account for the sequence $(5,3,1,1)$ since it is the only sequence which has its first three components identical to $\boldsymbol{a}$ and has a smaller index.
Thus we add $T_4^{36}=1$ to find the total index which is $70$.
\end{example}

The indexing algorithm (shaping) and its inverse (deshaping) can be implemented in a recursive way as $N$-step operations, see Algorithms 1 and 2 in \cite{gultekin4}.

\section{Simulation Setup and Performance metrics}\label{sec:simus}

\subsection{Simulation Setup }\label{sec:simu}
\begin{figure*}[tbp]
\centering		
\scalebox{0.8}{ \includegraphics{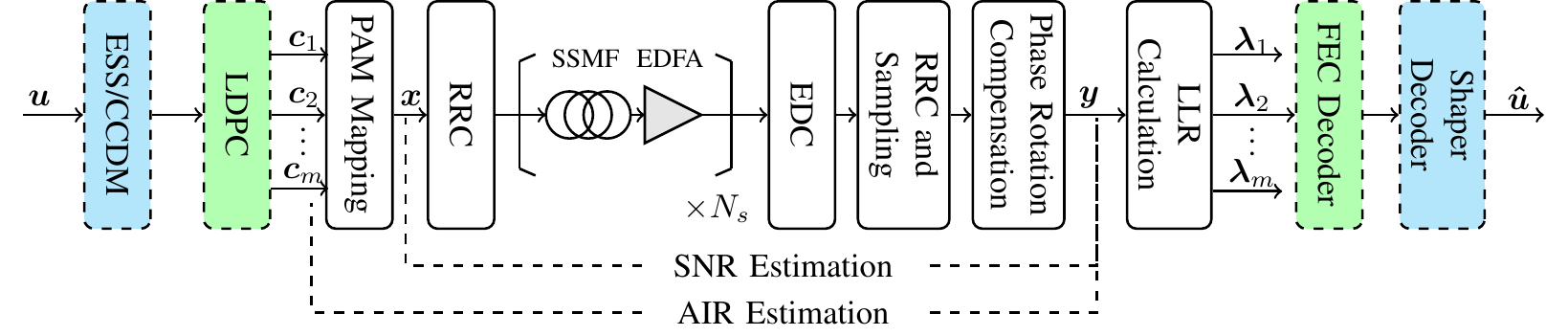} }
\caption{Transmission diagram: 
ESS: Enumerative sphere shaping, CCDM: Constant composition distribution matching, LDPC: Low-density parity check, PAM: Pulse amplitude modulation, RRC: Root-raised cosine, SSMF: standard single mode fiber, EDFA: Erbium-doped fiber amplifier, EDC: Electronic dispersion compensation, SNR: Signal-to-noise ratio, AIR: Achievable information rate, $N_s$: number of spans, $\lambda_1, \cdots,\lambda_m$: Log-likelihood ratios.}
\label{fig:6}
\end{figure*}

The performance of the ESS algorithm, which is implemented following \cite{gultekin2}, is compared to state-of-the-art CCDM-based probabilistic shaping and  uniform signaling through numerical simulations. Single-polarization single-channel and dual-polarization wavelength-division multiplexing (WDM) long-haul transmission configurations are considered. The effects of polarization mode dispersion and linear phase noise are neglected.

The simulation parameters for the WDM transmission are given by Table~\ref{tab:table2}. The net bit rate per WDM channel is \mbox{$405$ Gb/s}. 64-QAM (8-PAM per real dimension) is used as modulation formats for both shaped and uniform signaling. We consider low-density parity check (LDPC) codes with coding rates $R_c=3/4$ and $R_c=4/5$, as described in Example~\ref{example1}. This ensures two transmission scenarios (uniform and shaped) with same net bit rate. The shaping and coding parameters are given by Table~\ref{tab:table3}. It is important to note at this point that CCDM outputs amplitude sequences satisfying a given distribution, while ESS outputs sequences satisfying an energy constraint. Therefore, ESS cannot directly control the induced symbol distribution nor the corresponding entropy. Furthermore, the shaping rate is obtained differently in CCDM than in ESS. In the former, the input distribution needs to be adjusted, while in the latter, the maximum energy needs to be chosen. 

The transmission link consists of multi-span standard single mode fiber (SSMF) with an attenuation coefficient $\alpha=0.2~\mathrm{dB\cdot km^{-1}}$, a dispersion parameter $D=17 ~\mathrm{ps \cdot nm^{-1} \cdot km^{-1}}$, and a nonlinear coefficient \mbox{$\gamma=1.3~ \mathrm{W^{-1} \cdot km^{-1}}$}. An erbium-doped fiber amplifier (EDFA) with a $5$ dB noise figure and $16$ dB gain is used to periodically amplify the signal after each span of $L=80$~km.  

Each real-dimension of the noununiform signaling simulation setup is shown in Fig.~\ref{fig:6}. Firstly the information bits $\boldsymbol{u}$ are shaped using ESS or CCDM. The shaping and LDPC blocks operate as explained in Sec.~\ref{sec:SHPM} (see Fig.~\ref{fig:PASblockdiag}). After PAM mapping, the signal is oversampled with $16$ samples/symbol and passed through a root-raised cosine (RRC) filter, with roll-off factor $\rho = 0.1$, for spectrum shaping.
\begin{table}[tbp]
\centering
\caption{WDM simulation parameters.} \label{tab:table2}
\begin{tabular}[width=20mm]{cc}
\hline  Number of WDM channels  & 11 \\
\hline  Number of polarizations  & 2 \\
\hline  Symbol rate  & $45$ GBd\\
\hline  Channel spacing & $50$ GHz\\
\hline  Modulation  & $64$ QAM\\
\hline  RRC roll off $\rho$ & $0.1$\\
\hline  Span length $L$  & $80$ km\\
\hline  EDFA noise figure  & $5$ dB \\
\hline  Number of LDPC blocks & $100$\\
\hline
\end{tabular}
\end{table}

\begin{table}[tbp]
\centering
\caption{Shaping and coding parameters.} \label{tab:table3}
\scalebox{0.7}{
\begin{tabular}[width=20mm] {ccccc}
Name &  Uniform & CCDM-200 &  CCDM-3600 &ESS-200\\
\hline Bits per PAM symbol $m$      & 3 & 3    & 3   & 3 \\
\hline Block length $N$      & -             & 200    & 3600   & 200 \\
\hline Shaping rate $R_s$ [bits/amp.]     & - & 1.85 & 1.85 &  1.85   \\
\hline Entropy $H(P_A)$ [bits/amp.]    & - & 1.91 & 1.85   & 1.87 \\
\hline Rate loss $R_L$ [bits/amp.] & - & 0.06 & 0 & 0.02 \\
\hline FEC rate      & 3/4     & 4/5 & 4/5   & 4/5   \\
\hline Information rate $R$ [bits/1D-sym] &2.25 &2.25 &2.25 &2.25\\
\hline \\
\end{tabular}}
\end{table}

At the receiver side, after channel selection and downsampling the signal to $2$ samples/symbol, chromatic dispersion compensation is performed. After that, an RRC matched filter is applied before downsampling to $1$ sample/symbol. Phase rotation due to nonlinear effects is ideally compensated. At this stage, the effective SNR and BMD rate are estimated (see next section). Then, the log-likelihood ratios (LLRs) $\lambda_1, \lambda_2, \ldots,\lambda_m$ are calculated for the bits $c_1,c_2,\ldots,c_m$, which are then passed to the soft-decision LDPC decoder. Finally, ESS or CCDM deshaping is performed to recover the transmitted data bits. In the case of uniform signaling, the shaping and deshaping blocks are removed from the simulation setup (see Fig.~\ref{fig:PASblockdiag}~(a)). 

\subsection{Performance metrics}\label{sec:PM}
In this work, we mainly use two metrics to evaluate the performance of ESS in comparison with CCDM and uniform signaling: effective SNR, and bit-metric decoding (BMD) rate.

We follow standard nomenclature used in the literature to define the effective SNR as \cite{ps3} 
\begin{align}\label{eff.snr}
\text{SNR}_{\text{eff}} \approx \frac{\mathbb{E}[|{X}|^2]}{\mathbb{E}[|{Y}-{X}|^2]} 
\end{align}
where $\mathbb{E}[\cdot]$ represents expectation and $X$ and $Y$ are the transmitted and received symbols respectively. The effective SNR in \eqref{eff.snr} includes the amplified spontaneous emission (ASE) noise and the nonlinear noise, takes into accounts the probability of each constellation points. The effective SNR was calculated per QAM symbol.

The second metric we use is the ``finite blocklength BMD rate'' which was defined per QAM symbol ($2$D-sym) by Fehenberger {\it et al.} in~\cite[eq. (15)]{mp} as
\begin{equation}\label{mi}
\text{AIR}_{\text{N}} = 2\underbrace{\left[  H(\boldsymbol{C}) - \sum_{i=1}^m H(C_i \mid Y) \right]}_{\text{BMD Rate}} - 2 R_L,
\end{equation}
where $H(\cdot)$ denotes entropy. 
It is important to note that by using the more general framework discussed in \cite{R2}, the achievable rate in \eqref{mi} can be recovered in a straightforward manner\footnote{The achievable rate expression~\eqref{mi} can be derived from~\cite[eq. (1)]{R2} as
\begin{align}
\text{AIR}_{\text{N}} &\stackrel{(a)}{=} 2\left[ \frac{\log_2 \left| \calA^\bullet\right|}{N} + 1 - \mathbb{U}\left(q_{\text{bmd}},X,Y\right)  \right]^{+}, \nonumber \\
&\stackrel{(b)}{=} 2\left[ H(X)  - \sum_{j=1}^{m} H(C_j|Y) - R_L \right]^{+}, \label{eq:uncertainty2AIRn}
\end{align}
where $\mathbb{U}\left(q_{\text{bmd}},X,Y\right)$ is the {\it uncertainty} arising from the bitwise decoding metric $q_{\text{bmd}}$ as in~\cite[eq. (79)-80)]{R2}.
In~\eqref{eq:uncertainty2AIRn}, (a) follows from translating~\cite[eq. (1)]{R2} into the ESS context, and (b) is due to~\eqref{eq:shapingrate} and~\eqref{eq:rloss}. We note here that $H(X) = H(A) + 1$.}
We computed the BMD rate term in \eqref{mi} as in \cite{ps2}.
However, we also numerically checked that the same results can be obtained using the methodology explained in~\cite[Sec. VII-E]{R2}.
The definition of finite length BMD rate allows us to have fair comparisons for short blocklengths, where the rate loss could be significant. The finite length BMD rate in \eqref{mi} converges to the classic BMD rate for infinite blocklengths.

\section{Simulation Results: Single-polarization single-channel transmission}\label{sec:simr.1}

\begin{figure}[t]
\centering		
\resizebox{\columnwidth}{!}{ \includegraphics{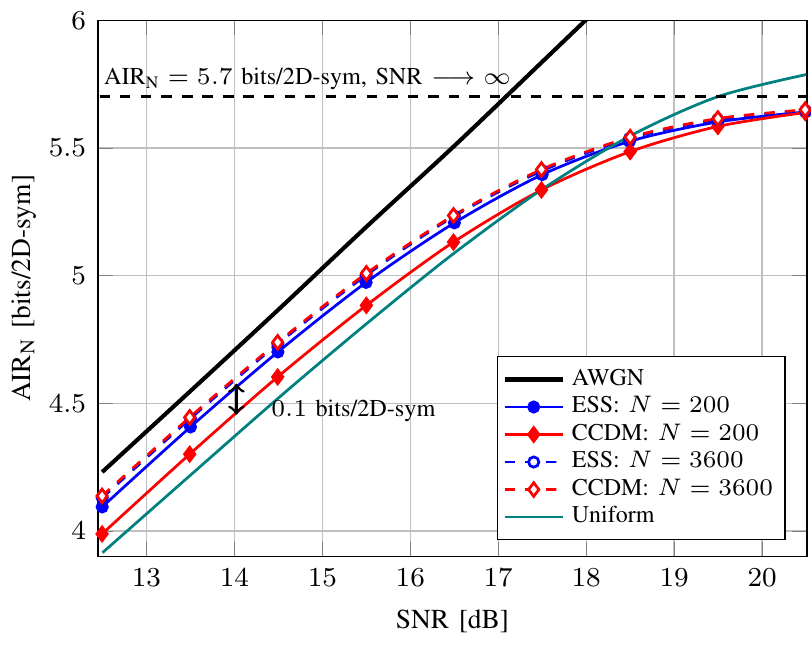} }
\caption{$\text{AIR}_{\text{N}}$ vs. SNR for linear fiber channel and different shaping blocklengths.}	
\label{fig:7}
\end{figure}

Here evaluate the performance of ESS and CCDM and uniform signaling in the context of single-polarization single-channel transmission.  To understand the effect of the shaping blocklength in the nonlinear fiber channel, we start by studying the linear channel, in which the fiber nonlinearity is turned off. In Fig.~\ref{fig:7}, we plot the BMD rate versus the SNR for different shaping blocklengths $N$. ESS performs better than CCDM at short blocklengths due to its lower rate loss. The gain for a blocklength $N=200$ is about $0.1$ bits/$2$D-sym at an SNR of $14$ dB. ESS and CCDM at $N=3600$ exhibit similar performances and show a gain of about $0.03$ bits/$2$D-sym and $0.19$ bits/$2$D-sym in comparison with ESS at $N=200$ and uniform, respectively at $14$ dB SNR. Increasing the blocklength in the linear channel improves the performance and reduces the gap with the AWGN capacity. For high SNR regime, shaped schemes converge to the maximum achievable ($\text{AIR}_\text{N} = 5.7$ bits/$2$D-sym) and the uniform signaling gives better performance.
 
\begin{figure}[t]
 \centering		
\resizebox{\columnwidth}{!}{ \includegraphics{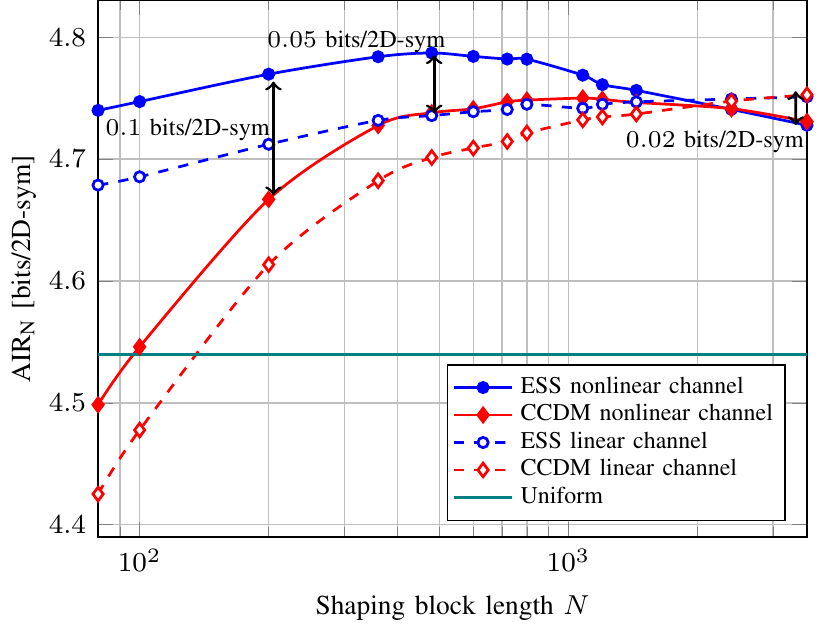} }
\caption{$\text{AIR}_{\text{N}}$ vs. shaping blocklength $N$  for linear and nonlinear fiber channel at optimal input power.}	
\label{fig:8}
\end{figure}

In Fig.~\ref{fig:8}, we evaluate the BMD rate as a function of the blocklength $N$ for both the linear and nonlinear channels. We first show the performance for the nonlinear fiber channel at optimal input power. Then, we fix the SNR for the linear channel so that the BMD rate performance for uniform signaling in linear and nonlinear channels are the same. In this case, in addition to the comparison of ESS, CCDM and uniform signaling in the nonlinear channel, we can also compare the gain/penalty of shaped schemes due to the fiber nonlinearity. 
As shown in  Fig.~\ref{fig:8}, for the linear channel case (dashed lines), ESS outperforms CCDM at short blocklengths due to its low rate loss. Then, at long blocklength ($N=3600$) in which the rate losses became negligible, ESS and CCDM converges to their maximum BMD rate and exhibits similar performances.

In the nonlinear channel case (solid lines in Fig.~\ref{fig:8}), we observe that in comparison with the linear channel, ESS and CCDM reach their maximum performance in terms of BMD rate at $N = 480$ and $N = 1200$, respectively. In addition, the ESS performance at $N=480$ increases by $0.05$ bits/$2$D-sym in comparison with linear channel performance. ESS still outperforms CCDM at short blocklength and the gain is $0.1$ bits/$2$D-sym at $N=200$. At long blocklengths, the performances of ESS and CCDM are slightly decreased unlike the linear channel transmission case. These results can be explained by the effective SNR performance, which measures the nonlinear tolerance of the different constellations.

\begin{figure}[t]
\centering		
\scalebox{1}{ \includegraphics{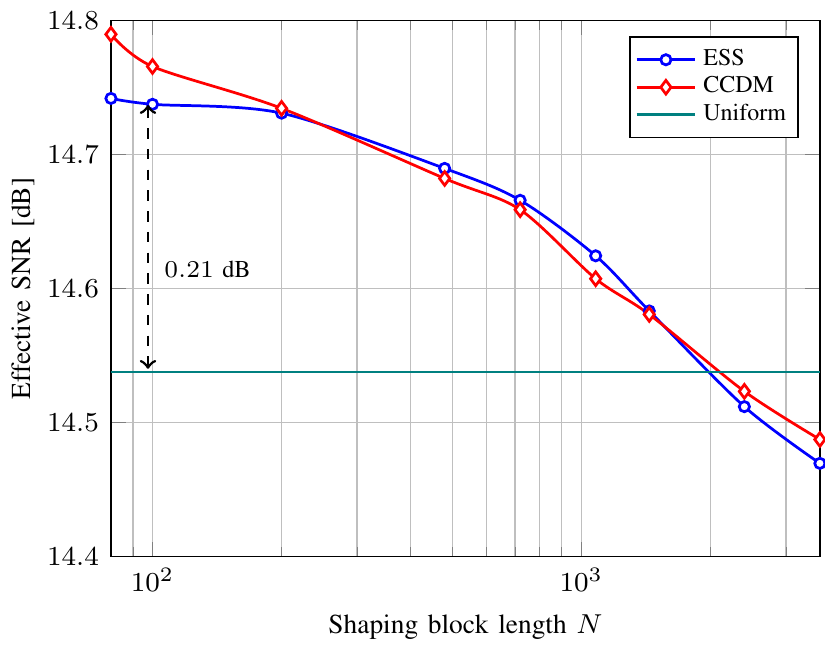} }
\caption{Effective SNR vs. shaping blocklength $N$  for nonlinear fiber channel at optimal input power. The effecive SNR gap between ESS and uniform at $N=100$ is shown.}	
\label{fig:9}
\end{figure}

In Fig.~\ref{fig:9}, we plot the effective SNR versus the shaping blocklength $N$ for nonlinear fiber channel at optimal input power. We observe that short blocklengths ESS and CCDM have better performance in terms of the effective SNR, and consequently higher nonlinear tolerance, than long blocklengths case. The gain for ESS at $N=100$ is about $0.27$ dB in comparison with ESS at $N=3600$. ESS and CCDM at the same blocklength exhibit similar performance in terms of effective SNR, and their performances are inversely proportional to the blocklength.

An important observation from Fig.~\ref{fig:9} is the fact that ESS and CCDM at short blocklengths give better SNR performance with respect to uniform signaling. Then, this performance gain is decreased for longer blocklengths. For blocklengths $N > 2000$, uniform signaling gives better performance than CCDM, which coincides with the state of the art results \cite{ps2}. The same behavior is observed with ESS. 
Our results therefore show that short-blocklength ESS (which exhibits the best performance) and CCDM, provide a combination of linear shaping gain and nonlinear tolerance gain in comparison with uniform. Long-blocklength ESS and CCDM provide higher linear shaping gain but at the same time, they are affected by a nonlinear penalty. The overall gain is lower than the short blocklength shaping schemes, and this, finite short-blocklength shaping is the best alternative.

\section{Simulation Results: Dual-polarization WDM transmission}\label{sec:simr.2}

In the following, we focus on the dual-polarization WDM configuration for which the simulation parameters are shown in Table~\ref{tab:table2}. We compare ESS and CCDM at a blocklength of $N = 200$, as well as CCDM with long blocklength ($N = 3600$), and uniform signaling.

\begin{figure}[t]
\centering		
\resizebox{\columnwidth}{!}{ \includegraphics{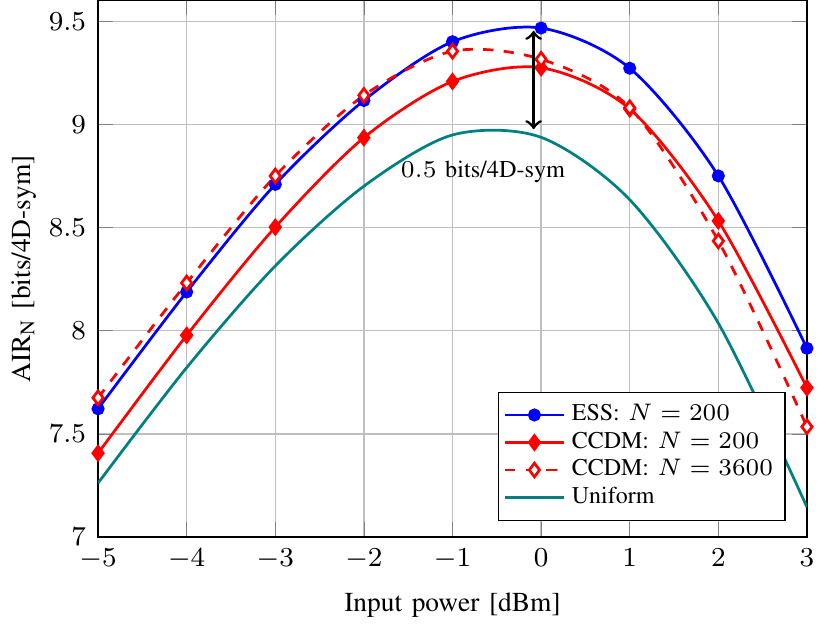} }
\caption{$\text{AIR}_{\text{N}}$ vs. input power for different blocklengths. At optimal input power, ESS at $N=200$ shows the highest BMD rate performance in comparison with CCDM at $N=200$ and $N=3600$, and uniform.}	
\label{fig:10}
\end{figure}

In Fig.~\ref{fig:10}, the BMD rate is plotted as a function of input power.
At optimal input power $0$ dBm, ESS at $N=200$ shows a gain of about $0.1$ bits/$4$D-sym, $0.19$ bits/$4$D-sym and $0.48$ bits/$4$D-sym in comparison with CCDM at $N=3600$, CCDM at $N=200$, and uniform signaling, respectively.
In the linear regime at low input powers, CCDM at $N=3600$ exhibits the highest linear shaping gain because it has the lowest rate loss. ESS at $N=200$ shows similar BMD rate performance to CCDM at $N=3600$ and better performance than CCDM at $N=200$. In the nonlinear regime, ESS and CCDM at $N=200$ show higher BMD rate than CCDM at $N=3600$. The performance gap between uniform signaling and CCDM at $N=3600$ is also reduced. This can be explained by the nonlinear penalty that long blocklengths CCDM suffers from, in comparison with uniform signaling and the short blocklengths case. 

\begin{figure}[t]
\centering		
\resizebox{\columnwidth}{!}{ \includegraphics{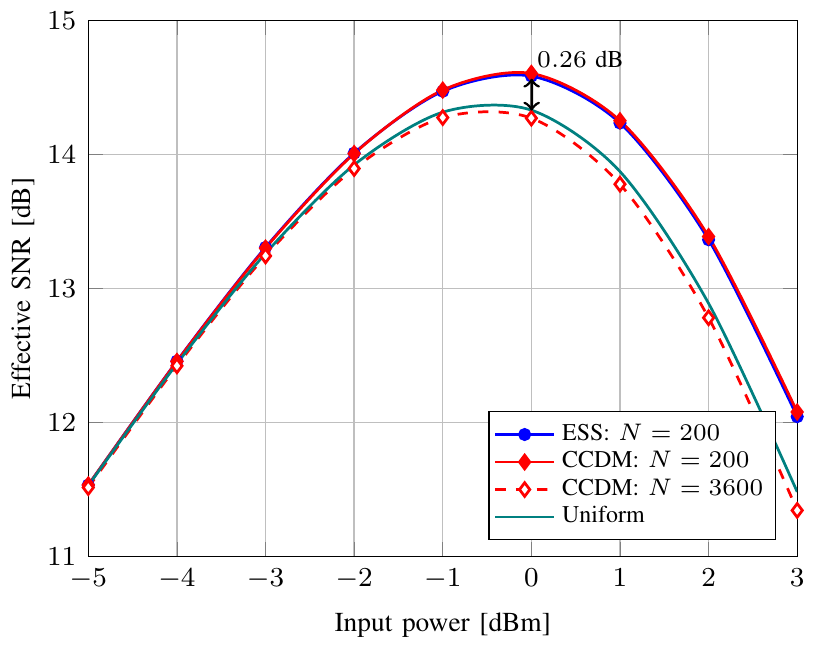} }
\caption{SNR vs. input power for different blocklengths. Short blocklengths ESS and CCDM exhibit higher nonlinear tolerance than uniform signaling and long blocklength CCDM. }	
\label{fig:11}
\end{figure}

The results in Fig.~\ref{fig:11}, in which the effective SNR is plotted as a function of the input power, confirms that ESS and CCDM at $N=200$ are more tolerant to the fiber nonlinearity. The gain at optimal input power is about $0.32$ dB and $0.26$ dB in comparison with long blocklength CCDM and uniform, respectively.  In the linear regime at low input powers, ESS, CCDM, and uniform signaling exhibits similar performance as expected for the case of linear channel.

\begin{figure}[t]
\centering		
\resizebox{\columnwidth}{!}{ \includegraphics{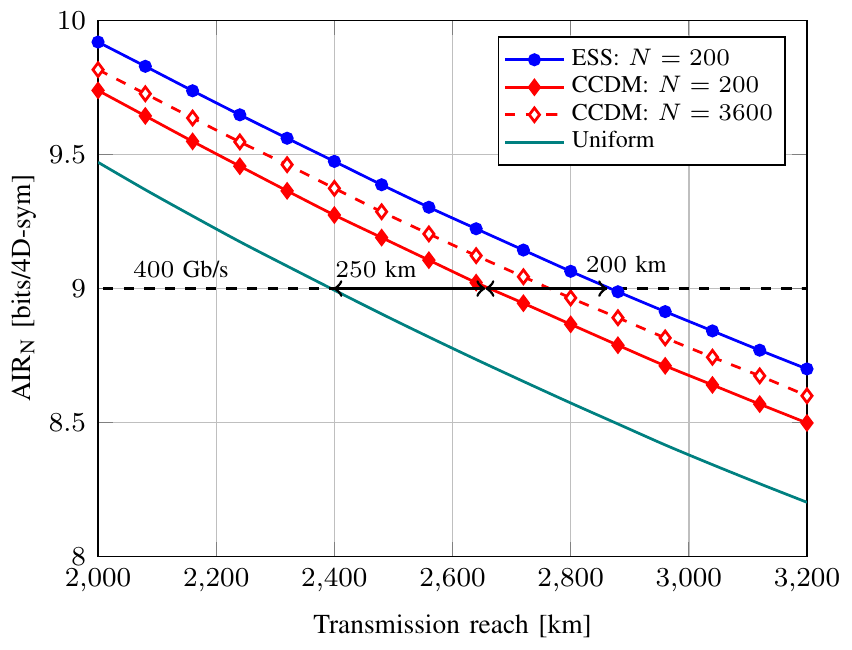} }
\caption{$\text{AIR}_{\text{N}}$ vs. transmission reach for different blocklengths at optimal input power.} 
\label{fig:12}
\end{figure}

We also evaluate the transmission reach increase obtained by using ESS and CCDM wth respect to the uniform signaling baseline. In Fig.~\ref{fig:12}, we plot BMD rate versus transmission reach  at optimal input power. At the net bit rate of $400$ Gb/s, it is observed that ESS at $N=200$ outperforms CCDM at $N=3600$, CCDM at $N=200$, and uniform signaling in terms of the transmission reach. The gains are about $100$ km, $200$ km  and $450$ km, respectively. This gain, in comparison with CCDM at $N=200$, is due to the linear shaping gain at short blocklength that ESS offers (see Figs.~\ref{fig:9}--\ref{fig:11}). For the uniform signaling case, ESS at $N=200$ provides a combination of linear shaping gain and nonlinear tolerance. Long blocklength CCDM with $N=3600$ exhibits lower performance in terms of transmission reach than ESS at $N=200$ due to the nonlinear penalty. CCDM at $N=200$ shows lower performance than CCDM at $N=3600$. This means that the large rate loss that CCDM at $N=200$ presents in comparison with $N=3600$, compensates the nonlinear gain provided by the use of short blocklength. 

\begin{figure}[t]
\centering		
\resizebox{\columnwidth}{!}{ \includegraphics{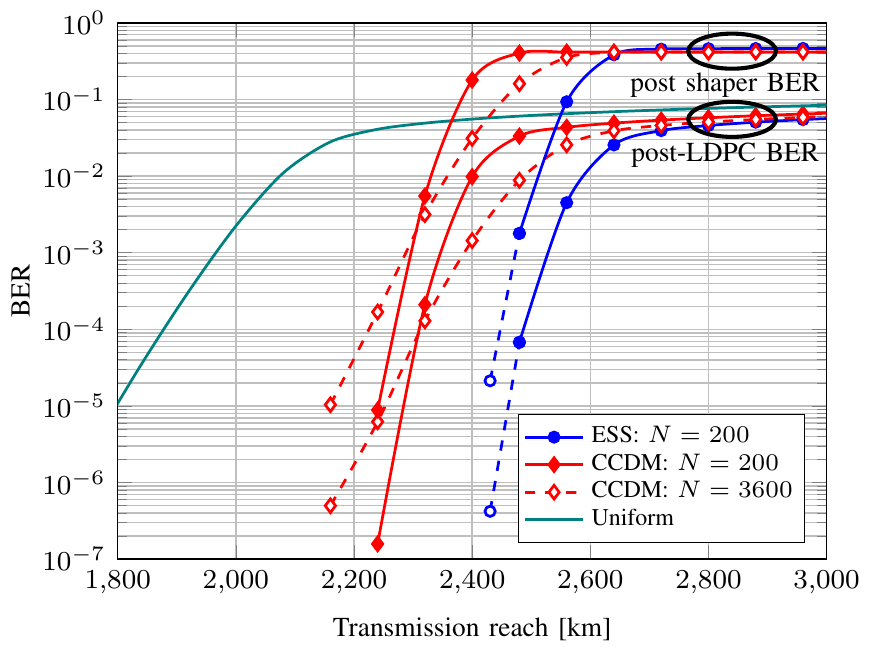} }
\caption{BER post-LDPC and BER post shaper vs. transmission reach for different blocklengths at optimal input power. For ESS, the dashed blue line is obtained by noise loading.}	
\label{fig:13}
\end{figure}

The transmission reach is also plotted as a function of the BER post-LDPC and BER post shaper, as given in Fig.~\ref{fig:13}. For each data point, 20000 LDPC blocks are simulated at the optimal launch power.
Noise loading is used to obtain data in the lower BER regime for ESS (indicated by dashed blue line and white filled marker).
The performance improvement of ESS at $N=200$ in terms of BMD rate are confirmed by the BER results. In addition, it is observed that the BER slopes in Fig.~\ref{fig:13} are different between methods. The slopes of CCDM with $N=200$ are similar to ESS with $N=200$, and the slopes of CCDM at $N=3600$ closely match that of uniform signaling. This difference in slopes can be explained by the effective SNR results shown in Fig.~\ref{fig:11}, where ESS and CCDM with $N=200$ exhibit similar performance, and CCDM with $N=3600$ shows slightly lower effective SNR than uniform.

\section{Conclusions}\label{sec:con}
We have proposed to use enumerative sphere shaping to increase to capacity of fiber optical communication systems. In the context of $200$ Gb/s 64-QAM single-polarization single channel and $400$ Gb/s 64-QAM dual-polarization WDM systems, we have shown that ESS improves the performance for short blocklengths in comparison with CCDM, due to its low rate loss. Short blocklengths ESS was shown to provide the best performance in terms of generalized mutual information and transmission reach. Short blocklengths ESS also exhibit better effective SNR performance than uniform signaling and long blocklengths ESS and CCDM. It combines both linear shaping gain and nonlinear tolerance, and has lower complexity than long blocklengths ESS and CCDM, which make it a promising candidate to be implemented in real-time systems. An experimental validation of ESS shaping and an investigation of the effective SNR improvement allowed by using short blocklength shaping is left for further investigation.

\section{Acknowledgements}
This work was supported by the Netherlands Organization for Scientific Research (NWO) via the VIDI Grant ICONIC (project number 15685). The work of A. Alvarado has received funding from the European Research Council (ERC) under the European Union's Horizon 2020 research and innovation programme (grant agreement No 757791). The authors would like to thank Dr. Tobias Fehenberger for fruitful discussions about probabilistic shaping.

\end{document}